\documentclass[11pt,letterpaper]{article}
\usepackage[T1]{fontenc}
\usepackage[utf8]{inputenc}
\usepackage{lmodern}
\usepackage[margin=1in]{geometry}
\usepackage{amsmath,amssymb,booktabs,array}
\usepackage{setspace,microtype}
\usepackage[round,authoryear]{natbib}
\usepackage{graphicx,pgfplots}
\pgfplotsset{compat=1.18}
\usepgfplotslibrary{fillbetween}
\usepackage{caption,placeins}
\usepackage[hidelinks]{hyperref}
\newcommand{\notes}[1]{\par\vspace{3pt}\begin{minipage}{\linewidth}\footnotesize\setstretch{1.04}\textit{Notes:} #1\end{minipage}}
\title{\vspace{-1.2cm}Extracurricular Exercise and Academic Achievement:\\[.3em] Evidence from the China Education Panel Survey}
\author{Ningyi Li}
\date{October 13, 2026}
\begin{document}
\maketitle
\begin{abstract}
\noindent Does physical exercise enhance adolescent academic achievement, and is there an optimal daily duration? Linking two waves of the China Education Panel Survey ($N=8,247$), we examine this association using restricted cubic splines and classroom fixed effects. Prospective estimates without prior-score controls place non-exercising students 0.162 mean subject-z units below peers exercising one hour daily. Flexible baseline-achievement controls shrink this shortfall by 85.1 percent to 0.024. The exercise block is not jointly significant ($p=0.301$), and curvature tests fail to reject linearity ($p=0.659$). Subgroup estimates yield similar gender profiles, with stronger evidence of location variation in English. These findings establish the importance of academic history in observed exercise gradients and challenge prescriptive exercise targets derived from fitted observational turning points.
\end{abstract}

\vspace{.3em}

\thispagestyle{empty}\newpage\setcounter{page}{1}

\section{Introduction}

Adolescent time allocation involves a fundamental economic trade-off between physical activity and academic effort. While exercise can support concentration and physical health, it inevitably competes with coursework, sleep, and leisure. Several empirical studies posit an inverted-U relationship, wherein academic performance peaks at moderate daily exercise durations \citep[e.g.,][]{zhang2021,wangguo2022}. A critical yet unresolved empirical question is whether this purported optimum reflects a true behavioral return or merely pre-existing student heterogeneity. Students who engage in moderate physical activity may systematically differ in academic endowment, parental investments, and school environments.

China's competitive secondary education system provides a compelling setting to test this trade-off. The China Education Panel Survey (CEPS) pairs rich longitudinal records on extracurricular time use and household backgrounds with school-supplied examination scores in Chinese, mathematics, and English. By tracking a baseline cohort of seventh-graders from the 2013--2014 academic year to their autumn-2014 midterm examinations, we can link initial exercise habits to subsequent achievement while directly conditioning on the complete vector of baseline subject marks.

We analyze 8,247 students across 112 junior high schools. To isolate conditional exercise contrasts, our empirical strategy combines baseline classroom fixed effects with flexible spline controls for prior academic history. A central feature of adolescent time use is the substantial point mass at zero: 27.9 percent of students report no extracurricular exercise during the preceding week. Rather than imposing a restrictive polynomial across the entire support, we decompose exercise exposure into an extensive-margin indicator (any exercise versus none) and an intensive-margin restricted cubic spline over positive durations. This flexible formulation accommodates discrete level shifts at zero while allowing the continuous conditional duration curve to depart from a quadratic shape.

Our central empirical finding is that prior academic performance accounts for the bulk of the prospective exercise gradient. In a prospective model controlling for student and family backgrounds and classroom fixed effects, non-exercising students score 0.162 mean subject-z units lower than peers exercising one hour daily. Once we flexibly control for baseline subject scores on the exact same sample, this gap shrinks by 85.1 percent to $-0.024$ mean subject-z units. The four exercise terms are not jointly significant ($p = 0.301$). A formal test of spline curvature fails to reject linearity ($p = 0.659$), and a simultaneous confidence band over the integer-minute grid from 5 to 120 encompasses zero. While Chinese language achievement exhibits a modest negative zero-exercise gap, estimated profiles are similar across gender, with the strongest evidence of school-location variation in English.

This paper makes three primary contributions. First, it demonstrates how strongly prospective exercise--achievement associations change after controlling for past academic performance, documenting this attenuation within a common longitudinal sample. Second, it decouples the zero-exercise contrast from the curvature of positive exercise, testing whether an intermediate optimum persists under flexible functional forms. Third, it provides a unified evaluation of heterogeneity across core academic subjects, gender, and school locations under formal multiple-testing adjustments.

The remainder of the paper is organized as follows. Section 2 reviews the related theoretical and empirical literature. Section 3 details the CEPS data and variable construction. Section 4 outlines our econometric framework. Section 5 presents the main estimates, heterogeneity analyses, and sensitivity checks. Section 6 discusses behavioral interpretations and data limitations. Section 7 concludes with policy implications.

\section{Background and Related Evidence}

In a human-capital framework, physical activity combines potential cognitive benefits with the opportunity cost of scarce study time. Exercise can support attention and executive function, while longer sessions may displace homework, revision, or rest. These competing channels motivate a nonlinear relationship. Estimating a flexible duration profile is therefore central to evaluating whether the data identify an interior optimum.

The broader empirical literature offers divergent findings. Randomized trials evaluating structured physical activity programs report mixed results: \citet{davis2011} report a positive dose-response trend in mathematics achievement among sedentary overweight children, whereas the Active Smarter Kids trial finds no overall improvement in its primary academic outcomes \citep{resaland2016}. Observational studies evaluate naturalistic time allocations but combine exercise differences with differences in student backgrounds. Prior academic performance is a central adjustment variable: it summarizes earlier educational inputs and experiences that can shape both exercise habits and subsequent test performance.

Studies employing the CEPS have reported that adolescent academic performance follows an inverted-U curve with respect to exercise. For example, \citet{zhang2021} estimate hierarchical panel models with a quadratic exercise term using two CEPS waves and report a turning point at 45.6 minutes per day; \citet{wangguo2022} report a peak at 3.55 hours per week in cross-sectional baseline data. Quadratic specifications impose a symmetric shape around their turning point, whose location can be sensitive to sparse tails and the concentration of observations at zero. Testing whether the estimated curvature persists with flexible splines and longitudinal controls for past performance is thus central to evaluating these findings.

To overcome these functional form restrictions, we draw on the modeling literature for continuous exposures with a point mass at zero \citep{royston2010} and implement restricted cubic splines for positive durations \citep{harrell2015}. We benchmark all contrasts against an intuitive reference of 60 minutes per day. Furthermore, we interact the full exercise block with student and school characteristics to assess whether exercise--achievement associations differ across learning domains or institutional environments.

\section{Data and Measurement}

\subsection{Longitudinal sample construction}

We utilize the seventh-grade cohort of the CEPS, which surveyed 10,279 students across 112 junior high schools in the 2013--2014 academic year and recorded school-supplied examination scores from autumn 2014 in the follow-up release. Our target sample consists of students with complete baseline demographics, eligible exercise reports, and valid school-supplied test scores across Chinese, Mathematics, and English. Among the 830 students recorded as unsuccessful questionnaire follow-ups in wave 2, 142 have complete school-supplied examination records and satisfy all baseline criteria. Because our dependent variable uses school examination records, we retain these 142 students in our benchmark sample to use their observed outcomes; restricting the sample to successful questionnaire follow-ups yields similar estimates. As documented in Table~\ref{tab:flow}, the final analytic sample consists of 8,247 students across 221 baseline classrooms.

\begin{table}[!htbp]\centering
\caption{From the baseline cohort to the analytic sample}\label{tab:flow}
\begin{small}\setstretch{1.04}\setlength{\tabcolsep}{4pt}\begin{tabular}{p{11.7cm}r}\toprule
Step & Students \\ \midrule
Baseline grade-7 cohort & 10,279 \\
Missing baseline exercise components or any subject score & $-693$ \\
Complete baseline exposure and scores & 9,586 \\
Reported exercise above five hours per day & $-111$ \\
Eligible exposure range and baseline scores & 9,475 \\
Missing one or more background controls & $-589$ \\
Baseline-eligible cohort & 8,886 \\
Unavailable or out-of-range follow-up outcome & $-639$ \\
Main analytic sample & 8,247 \\
Of which: official follow-up status 2 & 142 \\
Baseline classes / schools & 221 / 112 \\
\bottomrule\end{tabular}\end{small}
\notes{Exclusions are sequential. The 471 new entrants are outside the baseline cohort. Status 2 denotes unsuccessful survey follow-up. The final two rows describe the analytic sample.}
\end{table}

\subsection{Exercise exposure}

Baseline extracurricular exercise is measured as the daily average hours of physical activity during the preceding week, calculated from separate items for weekdays ($w$) and weekend days ($e$):

\begin{equation}
T_{i0}=\frac{5(H_{wi}+M_{wi}/60)+2(H_{ei}+M_{ei}/60)}{7}.
\end{equation}

Here $H$ and $M$ denote reported hours and minutes, respectively. Our primary sample restricts daily exercise to between zero and five hours per day, testing alternative caps of three, four, and eight hours in sensitivity analyses. The wave-2 survey instrument altered the exercise question, switching from weekday and weekend durations during the preceding week to usual weekly frequency and minutes on an exercising day. Because changes in questionnaire design confound within-student differencing, we treat baseline exercise as a prospective exposure rather than attempting a student fixed-effects specification across non-comparable survey items.

\subsection{Academic achievement}

Academic performance is measured using midterm examination scores in Chinese, Mathematics, and English. Baseline marks were standardized by survey administrators within school-by-grade groups to a mean of 70 and a standard deviation of 10, which we convert to subject z scores ($Z_{ij,0}$) by subtracting 70 and dividing by 10 \citep{ceps2015}. In the follow-up wave, schools administered decentralized midterm examinations with heterogeneous nominal maxima. We standardize raw marks within each recorded school and subject group to mean zero and unit variance ($Z_{isj,1}$), defining our primary outcome $Y_{i1}$ as the composite unweighted average across all three subjects:

\begin{equation}
Z_{isj,1}=\frac{R_{isj,1}-\overline R_{sj,1}}{S_{sj,1}},
\qquad Y_{i1}=\frac{1}{3}\sum_{j=1}^{3}Z_{isj,1}.
\end{equation}

This normalization ensures that the dependent variable captures \emph{relative} academic performance within each student's school cohort, expressed in mean subject-$z$ units (sample standard deviation is 0.851). The primary outcome excludes above-maximum scores, which affect 32 records with all three marks. Alternative scoring rules and percentage-of-maximum outcomes are examined in the robustness analysis. Reference distributions use all valid follow-up subject records before analytic restrictions. Appendix~\ref{app:variables} details normalization and score validation.

\subsection{Covariates and subgroups}

Vector $X_{i0}$ includes baseline demographic and socioeconomic controls: age, gender, minority status, agricultural hukou, migrant status, boarding status, siblings, parental education categories, household economic status, and recalled sixth-grade subject difficulty. Baseline cognitive test scores enter extended specifications. Subgroup analyses examine gender and school location, the latter defined via the principal-reported three-tier geographic measure (city or county centers, urban and rural--urban fringes, and towns and rural areas) to distinguish school settings. Table~\ref{tab:descriptive} presents descriptive statistics for the analytic sample. Covariates are selected by their measurement timing and substantive role \citep{vanderweele2019}.

\begin{table}[!htbp]\centering
\caption{Main-sample descriptive statistics}\label{tab:descriptive}
\begin{small}\setstretch{1.04}\setlength{\tabcolsep}{4pt}\begin{tabular}{p{10.8cm}rr}\toprule
Variable & Mean / share & SD \\ \midrule
Baseline exercise (minutes/day) & 51.29 & 55.69 \\
Baseline relative achievement index & 0.06 & 0.82 \\
Follow-up relative achievement index & 0.05 & 0.85 \\
Birth-year-based age & 12.52 & 0.70 \\
Zero reported exercise (percent) & 27.88 & -- \\
Female (percent) & 48.32 & -- \\
Agricultural registration (percent) & 51.59 & -- \\
Boarding (percent) & 29.87 & -- \\
\bottomrule\end{tabular}\end{small}
\notes{N=8,247. Statistics are unweighted. Achievement indices average three subject z scores, using the normalization described in Section 3.}
\end{table}

\section{Empirical Strategy}

\subsection{Econometric specification}

To assess whether exercise predicts subsequent academic achievement conditional on prior performance, we estimate the following prospective model:

\begin{equation}\label{eq:main}
Y_{i1}=\alpha_{c(i),0}+\delta D_{i0}
 +D_{i0}\{s(T_{i0})-s(1)\}
 +\sum_{j=1}^{3}f_j(Z_{ij,0})+X_{i0}'\gamma+\varepsilon_i.
\end{equation}

where $\alpha_{c(i),0}$ denotes baseline classroom fixed effects, absorbing shared initial peer, teacher, and school environments. The indicator $D_{i0} = \mathbf{1}(T_{i0} > 0)$ captures any exercise participation, while $s(T_{i0})$ is a restricted cubic spline with knots placed at the 5th, 35th, 65th, and 95th percentiles of positive exercise (approximately 14.3, 42.9, 77.1, and 182.5 minutes per day). Subtracting $s(1)$ centers the duration component at 60 minutes, rendering $-\delta$ the direct conditional contrast between zero exercise and one hour daily.

Crucially, rather than assuming linear persistence, each baseline subject score enters through a three-knot restricted cubic spline $f_j(Z_{ij,0})$ (knots at the 10th, 50th, and 90th percentiles), allowing academic persistence to vary nonlinearly across the achievement distribution. The nonlinear score terms are highly significant jointly ($F(3, 111) = 9.79, p < 0.001$). Across five repetitions of five-fold cross-validation holding out entire schools, the RMSE for class-demeaned scores falls from 0.433649 with linear score controls to 0.432127 with splines, a 0.351 percent improvement. Exercise knots are calculated among baseline-eligible seventh-graders before restricting follow-up outcomes; baseline-score knots use the estimation sample. The primary, subject, and group models share these knots. Appendix~\ref{app:variables} gives the reference populations and sample rules.

\subsection{Comparison sequence and inference}

To assess the role of prior achievement, we hold the sample, covariates, and fixed effects constant while comparing three sequential specifications: a baseline cross-section, a prospective model without prior scores, and our full specification in equation~\eqref{eq:main}. Standard errors are clustered at the school level (112 clusters) with finite-sample CR1 adjustments.

We evaluate exercise associations through multiple statistical benchmarks. First, we conduct joint Wald $F$-tests for the complete four-term exercise block as well as tests of spline curvature. Second, to account for uncertainty across the duration grid, we construct a 95 percent simultaneous confidence band over the integer-minute grid from 5 to 120 using 20,000 school-level Rademacher score-multiplier draws. Finally, subgroup interaction models by gender and school location report raw and Holm-adjusted $p$-values \citep{holm1979}. Adjustments cover two composite interaction tests and, separately, six subject interaction tests. An eight-restriction cross-equation test compares the three subject profiles. Subgroup models share the other control slopes, score normalization, and knots. Pointwise intervals are reported for individual contrasts. School clustering follows \citet{cameron2015}. Appendix~\ref{app:inference} describes the spline basis, cross-subject covariance, and simultaneous band.

\section{Estimation Results}

\subsection{The role of prior academic achievement}

Table~\ref{tab:sequence} reports the paper's central empirical result: controlling for prior academic achievement reduces the zero-versus-one-hour gap by 85.1 percent. In the cross-sectional benchmark (Column 1), students reporting zero exercise score 0.168 mean subject-z units lower than those exercising one hour daily ($p < 0.001$). Estimating a prospective model on follow-up scores without prior-score controls (Column 2) yields a nearly identical gap of $-0.162$ mean subject-z units ($p < 0.001$), showing that the unadjusted gradient persists when exercise precedes the outcome.

In Column 3, adding flexible spline controls for baseline subject scores reduces the zero-versus-one-hour gap to $-0.024$ mean subject-z units (95\% CI: $[-0.0485, 0.0005]$). This represents an 85.1 percent attenuation on the identical sample. Concurrently, the adjusted $R^2$ jumps from 0.273 to 0.737, and a joint Wald test across all four exercise terms fails to reject the null hypothesis ($F = 1.23, p = 0.301$). These estimates establish the importance of academic history in the prospective exercise--achievement relationship.

\begin{table}[!htbp]\centering
\caption{Exercise contrasts across sequential specifications}\label{tab:sequence}
\begin{small}\setstretch{1.04}\setlength{\tabcolsep}{4pt}\begin{tabular}{lccc}\toprule
Contrast / statistic & \shortstack{Baseline outcome\\Background controls} & \shortstack{Follow-up outcome\\Background controls} & \shortstack{Follow-up outcome\\Plus baseline scores} \\ \midrule
0 vs. 60 min & \shortstack{-0.1682\\{\scriptsize [-0.2077, -0.1287]}} & \shortstack{-0.1618\\{\scriptsize [-0.2043, -0.1192]}} & \shortstack{-0.0240\\{\scriptsize [-0.0485, 0.0005]}} \\
30 vs. 60 min & \shortstack{0.0203\\{\scriptsize [-0.0049, 0.0455]}} & \shortstack{0.0149\\{\scriptsize [-0.0135, 0.0433]}} & \shortstack{-0.0026\\{\scriptsize [-0.0180, 0.0128]}} \\
120 vs. 60 min & \shortstack{-0.0889\\{\scriptsize [-0.1352, -0.0427]}} & \shortstack{-0.0915\\{\scriptsize [-0.1413, -0.0418]}} & \shortstack{-0.0160\\{\scriptsize [-0.0464, 0.0145]}} \\
Joint exercise-test $p$ & $<0.001$ & $<0.001$ & 0.301 \\
Adjusted $R^2$ & 0.281 & 0.273 & 0.737 \\
Students & 8,247 & 8,247 & 8,247 \\
\bottomrule\end{tabular}\end{small}
\notes{All columns use the same students, background controls, baseline class fixed effects, and school-clustered inference. Column 3 adds three-knot splines for each baseline subject score. Brackets report pointwise 95\% confidence intervals. Contrasts are in mean subject-z units.}
\end{table}

\subsection{Exercise duration and core academic subjects}

Table~\ref{tab:contrasts} and Figure~\ref{fig:curve} examine the conditional duration profile across positive exercise durations. Relative to the one-hour benchmark, point estimates at the reported durations are small: $-0.003$ mean subject-z units for 30 minutes and $-0.016$ mean subject-z units for 120 minutes. A formal test of spline curvature fails to reject a linear relationship ($F = 0.42, p = 0.659$). As shown in Figure~\ref{fig:curve}, the 95 percent simultaneous confidence band covers zero over the integer-minute grid from 5 to 120. The flexible profile does not identify a distinct interior optimum.

\begin{table}[!htbp]\centering
\caption{Primary prospective contrasts relative to 60 minutes per day}\label{tab:contrasts}
\begin{small}\setstretch{1.04}\setlength{\tabcolsep}{4pt}\begin{tabular}{rrrr}\toprule
Minutes/day & Estimate & Clustered SE & Pointwise 95\% CI \\ \midrule
0 & -0.0240 & 0.0124 & [-0.0485, 0.0005] \\
15 & -0.0114 & 0.0177 & [-0.0464, 0.0237] \\
30 & -0.0026 & 0.0078 & [-0.0180, 0.0128] \\
60 & 0.0000 & 0.0000 & [0.0000, 0.0000] \\
90 & -0.0098 & 0.0111 & [-0.0319, 0.0123] \\
120 & -0.0160 & 0.0154 & [-0.0464, 0.0145] \\
\bottomrule\end{tabular}\end{small}
\notes{N=8,247; 221 baseline classes; 112 schools. Estimates are in mean subject-z units. The 60-minute reference is zero by normalization.}
\end{table}

\begin{figure}[!htbp]\centering
\begin{tikzpicture}
\begin{axis}[width=.92\linewidth,height=7.5cm,xmin=-3,xmax=122,ymin=-.10,ymax=.075,
axis lines=left,xlabel={Baseline extracurricular exercise (minutes/day)},ylabel={Follow-up achievement contrast},
xtick={0,15,30,60,90,120},ytick={-.10,-.05,0,.05},tick label style={font=\small},label style={font=\small},
yticklabel style={/pgf/number format/fixed,/pgf/number format/precision=2},
legend columns=2,legend style={font=\footnotesize,draw=none,at={(.5,1.04)},anchor=south,fill=none}]
\addplot[name path=lower,draw=none,forget plot] coordinates {(5,-0.07982255) (6,-0.07726788) (7,-0.07471679) (8,-0.07216962) (9,-0.06962680) (10,-0.06708881) (11,-0.06455617) (12,-0.06202953) (13,-0.05950961) (14,-0.05699725) (15,-0.05449368) (16,-0.05200272) (17,-0.04952963) (18,-0.04707978) (19,-0.04465863) (20,-0.04227172) (21,-0.03992472) (22,-0.03762338) (23,-0.03537357) (24,-0.03318127) (25,-0.03105257) (26,-0.02899364) (27,-0.02701069) (28,-0.02510994) (29,-0.02329750) (30,-0.02157926) (31,-0.01996073) (32,-0.01844677) (33,-0.01704132) (34,-0.01574705) (35,-0.01456500) (36,-0.01349423) (37,-0.01253153) (38,-0.01167129) (39,-0.01090557) (40,-0.01022430) (41,-0.00961571) (42,-0.00906692) (43,-0.00856442) (44,-0.00809500) (45,-0.00764714) (46,-0.00721124) (47,-0.00677945) (48,-0.00634540) (49,-0.00590409) (50,-0.00545167) (51,-0.00498524) (52,-0.00450275) (53,-0.00400282) (54,-0.00348465) (55,-0.00294793) (56,-0.00239276) (57,-0.00181955) (58,-0.00122902) (59,-0.00062212) (60,0.00000000) (61,-0.00119662) (62,-0.00243182) (63,-0.00370167) (64,-0.00500214) (65,-0.00632909) (66,-0.00767833) (67,-0.00904560) (68,-0.01042657) (69,-0.01181688) (70,-0.01321212) (71,-0.01460785) (72,-0.01599959) (73,-0.01738284) (74,-0.01875309) (75,-0.02010581) (76,-0.02143643) (77,-0.02274040) (78,-0.02401367) (79,-0.02525535) (80,-0.02646575) (81,-0.02764520) (82,-0.02879404) (83,-0.02991262) (84,-0.03100128) (85,-0.03206035) (86,-0.03309021) (87,-0.03409120) (88,-0.03506367) (89,-0.03600799) (90,-0.03692451) (91,-0.03781361) (92,-0.03867564) (93,-0.03951098) (94,-0.04031999) (95,-0.04110305) (96,-0.04186052) (97,-0.04259280) (98,-0.04330024) (99,-0.04398325) (100,-0.04464219) (101,-0.04527746) (102,-0.04588944) (103,-0.04647853) (104,-0.04704511) (105,-0.04758959) (106,-0.04811236) (107,-0.04861382) (108,-0.04909438) (109,-0.04955445) (110,-0.04999443) (111,-0.05041474) (112,-0.05081579) (113,-0.05119801) (114,-0.05156182) (115,-0.05190765) (116,-0.05223592) (117,-0.05254707) (118,-0.05284155) (119,-0.05311978) (120,-0.05338221)};
\addplot[name path=upper,draw=none,forget plot] coordinates {(5,0.04441511) (6,0.04312754) (7,0.04184354) (8,0.04056348) (9,0.03928776) (10,0.03801686) (11,0.03675132) (12,0.03549178) (13,0.03423895) (14,0.03299368) (15,0.03175707) (16,0.03053133) (17,0.02931935) (18,0.02812408) (19,0.02694858) (20,0.02579600) (21,0.02466959) (22,0.02357270) (23,0.02250880) (24,0.02148147) (25,0.02049439) (26,0.01955132) (27,0.01865608) (28,0.01781248) (29,0.01702423) (30,0.01629482) (31,0.01562734) (32,0.01502426) (33,0.01448712) (34,0.01401618) (35,0.01361007) (36,0.01326546) (37,0.01297673) (38,0.01273587) (39,0.01253252) (40,0.01235423) (41,0.01218682) (42,0.01201499) (43,0.01182285) (44,0.01159594) (45,0.01132461) (46,0.01100160) (47,0.01062137) (48,0.01017993) (49,0.00967462) (50,0.00910392) (51,0.00846729) (52,0.00776500) (53,0.00699803) (54,0.00616792) (55,0.00527671) (56,0.00432682) (57,0.00332104) (58,0.00226241) (59,0.00115423) (60,0.00000000) (61,0.00063601) (62,0.00128445) (63,0.00194374) (64,0.00261218) (65,0.00328798) (66,0.00396931) (67,0.00465422) (68,0.00534076) (69,0.00602689) (70,0.00671055) (71,0.00738963) (72,0.00806202) (73,0.00872556) (74,0.00937806) (75,0.01001735) (76,0.01064120) (77,0.01124742) (78,0.01183402) (79,0.01240062) (80,0.01294742) (81,0.01347464) (82,0.01398251) (83,0.01447126) (84,0.01494111) (85,0.01539232) (86,0.01582512) (87,0.01623974) (88,0.01663645) (89,0.01701548) (90,0.01737709) (91,0.01772153) (92,0.01804904) (93,0.01835990) (94,0.01865436) (95,0.01893268) (96,0.01919512) (97,0.01944195) (98,0.01967344) (99,0.01988985) (100,0.02009147) (101,0.02027855) (102,0.02045139) (103,0.02061026) (104,0.02075544) (105,0.02088722) (106,0.02100588) (107,0.02111172) (108,0.02120503) (109,0.02128611) (110,0.02135525) (111,0.02141276) (112,0.02145894) (113,0.02149411) (114,0.02151857) (115,0.02153264) (116,0.02153664) (117,0.02153090) (118,0.02151574) (119,0.02149148) (120,0.02145848)};
\addplot[fill=gray!20,draw=none] fill between[of=lower and upper];\addlegendentry{Simultaneous 95\% band}
\addplot[name path=plower,draw=none,forget plot] coordinates {(5,-0.06821278) (6,-0.06601716) (7,-0.06382445) (8,-0.06163493) (9,-0.05944895) (10,-0.05726688) (11,-0.05508918) (12,-0.05291635) (13,-0.05074898) (14,-0.04858777) (15,-0.04643371) (16,-0.04429007) (17,-0.04216134) (18,-0.04005212) (19,-0.03796707) (20,-0.03591092) (21,-0.03388851) (22,-0.03190472) (23,-0.02996457) (24,-0.02807314) (25,-0.02623561) (26,-0.02445721) (27,-0.02274322) (28,-0.02109892) (29,-0.01952951) (30,-0.01804000) (31,-0.01663509) (32,-0.01531897) (33,-0.01409504) (34,-0.01296573) (35,-0.01193209) (36,-0.01099359) (37,-0.01014783) (38,-0.00939049) (39,-0.00871533) (40,-0.00811438) (41,-0.00757831) (42,-0.00709685) (43,-0.00665927) (44,-0.00625492) (45,-0.00587426) (46,-0.00550929) (47,-0.00515337) (48,-0.00480114) (49,-0.00444829) (50,-0.00409148) (51,-0.00372813) (52,-0.00335635) (53,-0.00297481) (54,-0.00258264) (55,-0.00217936) (56,-0.00176482) (57,-0.00133917) (58,-0.00090275) (59,-0.00045612) (60,0.00000000) (61,-0.00102537) (62,-0.00208454) (63,-0.00317412) (64,-0.00429059) (65,-0.00543039) (66,-0.00658988) (67,-0.00776538) (68,-0.00895315) (69,-0.01014942) (70,-0.01135039) (71,-0.01255223) (72,-0.01375108) (73,-0.01494306) (74,-0.01612429) (75,-0.01729085) (76,-0.01843884) (77,-0.01956431) (78,-0.02066377) (79,-0.02173647) (80,-0.02278266) (81,-0.02380263) (82,-0.02479666) (83,-0.02576503) (84,-0.02670805) (85,-0.02762599) (86,-0.02851917) (87,-0.02938787) (88,-0.03023239) (89,-0.03105305) (90,-0.03185013) (91,-0.03262396) (92,-0.03337483) (93,-0.03410306) (94,-0.03480895) (95,-0.03549282) (96,-0.03615499) (97,-0.03679577) (98,-0.03741547) (99,-0.03801443) (100,-0.03859296) (101,-0.03915138) (102,-0.03969002) (103,-0.04020921) (104,-0.04070928) (105,-0.04119056) (106,-0.04165339) (107,-0.04209810) (108,-0.04252504) (109,-0.04293453) (110,-0.04332694) (111,-0.04370260) (112,-0.04406186) (113,-0.04440507) (114,-0.04473260) (115,-0.04504479) (116,-0.04534202) (117,-0.04562463) (118,-0.04589300) (119,-0.04614750) (120,-0.04638849)};
\addplot[name path=pupper,draw=none,forget plot] coordinates {(5,0.03280534) (6,0.03187682) (7,0.03095120) (8,0.03002878) (9,0.02910990) (10,0.02819493) (11,0.02728433) (12,0.02637860) (13,0.02547833) (14,0.02458420) (15,0.02369710) (16,0.02281868) (17,0.02195106) (18,0.02109642) (19,0.02025703) (20,0.01943520) (21,0.01863337) (22,0.01785404) (23,0.01709980) (24,0.01637334) (25,0.01567742) (26,0.01501489) (27,0.01438861) (28,0.01380146) (29,0.01325624) (30,0.01275556) (31,0.01230170) (32,0.01189646) (33,0.01154085) (34,0.01123486) (35,0.01097717) (36,0.01076482) (37,0.01059303) (38,0.01045507) (39,0.01034228) (40,0.01024431) (41,0.01014941) (42,0.01004492) (43,0.00991770) (44,0.00975586) (45,0.00955173) (46,0.00929964) (47,0.00899530) (48,0.00863567) (49,0.00821882) (50,0.00774373) (51,0.00721017) (52,0.00661860) (53,0.00597002) (54,0.00526590) (55,0.00450813) (56,0.00369889) (57,0.00284066) (58,0.00193614) (59,0.00098824) (60,0.00000000) (61,0.00046476) (62,0.00093718) (63,0.00141619) (64,0.00190063) (65,0.00238929) (66,0.00288086) (67,0.00337400) (68,0.00386733) (69,0.00435942) (70,0.00484881) (71,0.00533401) (72,0.00581351) (73,0.00628578) (74,0.00674926) (75,0.00720239) (76,0.00764361) (77,0.00807132) (78,0.00848413) (79,0.00888174) (80,0.00926433) (81,0.00963207) (82,0.00998512) (83,0.01032367) (84,0.01064788) (85,0.01095796) (86,0.01125407) (87,0.01153641) (88,0.01180517) (89,0.01206054) (90,0.01230271) (91,0.01253187) (92,0.01274823) (93,0.01295198) (94,0.01314332) (95,0.01332246) (96,0.01348959) (97,0.01364492) (98,0.01378867) (99,0.01392104) (100,0.01404223) (101,0.01415247) (102,0.01425197) (103,0.01434094) (104,0.01441961) (105,0.01448819) (106,0.01454691) (107,0.01459600) (108,0.01463569) (109,0.01466619) (110,0.01468776) (111,0.01470062) (112,0.01470501) (113,0.01470117) (114,0.01468934) (115,0.01466978) (116,0.01464274) (117,0.01460845) (118,0.01456719) (119,0.01451920) (120,0.01446476)};
\addplot[fill=blue!20,draw=none] fill between[of=plower and pupper];\addlegendentry{Pointwise 95\% interval}
\addplot[thick,blue!65!black] coordinates {(5,-0.01770372) (6,-0.01707017) (7,-0.01643662) (8,-0.01580307) (9,-0.01516952) (10,-0.01453598) (11,-0.01390243) (12,-0.01326888) (13,-0.01263533) (14,-0.01200178) (15,-0.01136830) (16,-0.01073569) (17,-0.01010514) (18,-0.00947785) (19,-0.00885502) (20,-0.00823786) (21,-0.00762757) (22,-0.00702534) (23,-0.00643239) (24,-0.00584990) (25,-0.00527909) (26,-0.00472116) (27,-0.00417731) (28,-0.00364873) (29,-0.00313663) (30,-0.00264222) (31,-0.00216669) (32,-0.00171125) (33,-0.00127710) (34,-0.00086544) (35,-0.00047746) (36,-0.00011438) (37,0.00022260) (38,0.00053229) (39,0.00081348) (40,0.00106496) (41,0.00128555) (42,0.00147403) (43,0.00162921) (44,0.00175047) (45,0.00183874) (46,0.00189518) (47,0.00192096) (48,0.00191727) (49,0.00188527) (50,0.00182613) (51,0.00174102) (52,0.00163112) (53,0.00149760) (54,0.00134163) (55,0.00116439) (56,0.00096703) (57,0.00075075) (58,0.00051670) (59,0.00026606) (60,0.00000000) (61,-0.00028030) (62,-0.00057368) (63,-0.00087897) (64,-0.00119498) (65,-0.00152055) (66,-0.00185451) (67,-0.00219569) (68,-0.00254291) (69,-0.00289500) (70,-0.00325079) (71,-0.00360911) (72,-0.00396878) (73,-0.00432864) (74,-0.00468752) (75,-0.00504423) (76,-0.00539761) (77,-0.00574649) (78,-0.00608982) (79,-0.00642736) (80,-0.00675916) (81,-0.00708528) (82,-0.00740577) (83,-0.00772068) (84,-0.00803008) (85,-0.00833402) (86,-0.00863255) (87,-0.00892573) (88,-0.00921361) (89,-0.00949625) (90,-0.00977371) (91,-0.01004604) (92,-0.01031330) (93,-0.01057554) (94,-0.01083281) (95,-0.01108518) (96,-0.01133270) (97,-0.01157542) (98,-0.01181340) (99,-0.01204670) (100,-0.01227536) (101,-0.01249945) (102,-0.01271903) (103,-0.01293414) (104,-0.01314484) (105,-0.01335119) (106,-0.01355324) (107,-0.01375105) (108,-0.01394468) (109,-0.01413417) (110,-0.01431959) (111,-0.01450099) (112,-0.01467843) (113,-0.01485195) (114,-0.01502163) (115,-0.01518751) (116,-0.01534964) (117,-0.01550809) (118,-0.01566290) (119,-0.01581415) (120,-0.01596187)};\addlegendentry{Positive-duration contrast}
\addplot[only marks,black,mark=*,error bars/.cd,y dir=both,y explicit] coordinates {(0,-0.024042187) +- (0,0.024495784)};
\addlegendentry{Zero report: pointwise CI}
\addplot[gray,dashed,forget plot] coordinates {(-3,0)(122,0)};
\end{axis}\end{tikzpicture}
\caption{Baseline exercise and subsequent achievement, conditional on baseline scores}\label{fig:curve}
\notes{N=8,247. Contrasts are relative to 60 minutes/day in mean subject-z units. The simultaneous band uses 20,000 school-level Rademacher score-multiplier draws over the integer-minute grid from 5 through 120. The zero-exercise point has a separate pointwise interval.}
\end{figure}

\FloatBarrier

Table~\ref{tab:subjects} presents subject-specific estimates. The zero-exercise gap is modestly larger for Chinese ($-0.045$ subject-z units, 95\% CI: $[-0.080, -0.010]$) than for mathematics ($-0.026$ subject-z units) or English ($-0.001$ subject-z units). However, the exercise block is not jointly significant in any individual subject (all $p > 0.14$), and a joint test across all eight cross-equation restrictions fails to reject coefficient equality across subjects ($p = 0.453$).

\begin{table}[!htbp]\centering
\caption{Subject-specific follow-up outcomes}\label{tab:subjects}
\begin{footnotesize}\setstretch{1.04}\setlength{\tabcolsep}{4pt}\begin{tabular}{p{3.8cm}rccc r}\toprule
Specification & N & 0 vs. 60 & 30 vs. 60 & 120 vs. 60 & Joint $p$ \\ \midrule
Chinese & 8,247 & \shortstack{-0.0452\\{\scriptsize [-0.0804, -0.0101]}} & \shortstack{-0.0054\\{\scriptsize [-0.0266, 0.0158]}} & \shortstack{-0.0204\\{\scriptsize [-0.0605, 0.0196]}} & 0.145 \\
Mathematics & 8,247 & \shortstack{-0.0263\\{\scriptsize [-0.0618, 0.0091]}} & \shortstack{-0.0134\\{\scriptsize [-0.0370, 0.0103]}} & \shortstack{-0.0212\\{\scriptsize [-0.0627, 0.0204]}} & 0.551 \\
English & 8,247 & \shortstack{-0.0005\\{\scriptsize [-0.0336, 0.0325]}} & \shortstack{0.0108\\{\scriptsize [-0.0101, 0.0317]}} & \shortstack{-0.0063\\{\scriptsize [-0.0433, 0.0308]}} & 0.746 \\
\bottomrule\end{tabular}\end{footnotesize}
\notes{All rows use the same 8,247 students, background controls, baseline class effects, and three-knot splines for all three baseline subject scores. Units are subject z scores. Intervals are pointwise. Each joint test covers the four exercise coefficients for that subject.}
\end{table}

\subsection{Subgroup heterogeneity: gender and school location}

Tables~\ref{tab:gender} and~\ref{tab:location} evaluate heterogeneity across demographic and institutional subgroups. Estimated exercise--achievement profiles are similar across gender (Table~\ref{tab:gender}): the zero-exercise shortfall is $-0.028$ mean subject-z units for boys and $-0.021$ for girls, and a joint test of gender-by-exercise interactions does not reject equality ($p = 0.670$, Holm-adjusted $p = 0.670$).

\begin{table}[!htbp]\centering
\caption{Exercise and achievement by gender}\label{tab:gender}
\begin{footnotesize}\setstretch{1.04}\setlength{\tabcolsep}{4pt}\begin{tabular}{p{4.0cm}cccc}\toprule
Group / statistic & Composite & Chinese & Mathematics & English \\ \midrule
Boys ($N=4,262$) & \shortstack{-0.0277\\{\scriptsize [-0.0662, 0.0108]}} & \shortstack{-0.0422\\{\scriptsize [-0.1014, 0.0171]}} & \shortstack{-0.0429\\{\scriptsize [-0.0973, 0.0116]}} & \shortstack{0.0020\\{\scriptsize [-0.0453, 0.0493]}} \\
Girls ($N=3,985$) & \shortstack{-0.0214\\{\scriptsize [-0.0503, 0.0076]}} & \shortstack{-0.0482\\{\scriptsize [-0.0884, -0.0080]}} & \shortstack{-0.0125\\{\scriptsize [-0.0571, 0.0321]}} & \shortstack{-0.0035\\{\scriptsize [-0.0419, 0.0350]}} \\
Interaction test: raw $p$ & 0.670 & 0.686 & 0.482 & 0.546 \\
Interaction test: Holm $p$ & 0.670 & 1.000 & 1.000 & 1.000 \\
\bottomrule\end{tabular}\end{footnotesize}
\notes{Entries compare zero exercise with 60 minutes/day; brackets give pointwise 95\% school-clustered intervals. Each outcome uses a pooled model on 8,247 students, interacting the full exercise block with group membership and sharing other control slopes. Baseline class effects and baseline-score splines match the primary model. Composite units are mean subject z scores; subject columns use subject z scores. Interaction tests have 4 numerator and 111 denominator degrees of freedom. Holm adjustments cover two composite interaction tests and, separately, six subject interaction tests.}
\end{table}

Geographic comparisons in Table~\ref{tab:location} reveal suggestive disparities. The zero-exercise gap is largest among students in city or county centers ($-0.055$ mean subject-z units, $p = 0.001$), whereas it is closer to zero in towns and rural areas ($-0.008$ mean subject-z units) and urban and rural--urban fringes ($0.011$ mean subject-z units). The omnibus location-interaction test yields $p=0.056$ before and $p=0.113$ after Holm correction. Evidence of location variation is strongest for English ($p=0.015$, Holm-adjusted $p=0.087$), identifying a useful focus for further research on school environments and extracurricular routines.

\begin{table}[!htbp]\centering
\caption{Exercise and achievement by school location}\label{tab:location}
\begin{footnotesize}\setstretch{1.04}\setlength{\tabcolsep}{4pt}\begin{tabular}{p{4.0cm}cccc}\toprule
Group / statistic & Composite & Chinese & Mathematics & English \\ \midrule
City or county center ($N=3,372$) & \shortstack{-0.0553\\{\scriptsize [-0.0863, -0.0244]}} & \shortstack{-0.0751\\{\scriptsize [-0.1284, -0.0218]}} & \shortstack{-0.0393\\{\scriptsize [-0.0963, 0.0177]}} & \shortstack{-0.0516\\{\scriptsize [-0.0943, -0.0090]}} \\
Urban / rural--urban fringe ($N=2,083$) & \shortstack{0.0113\\{\scriptsize [-0.0520, 0.0745]}} & \shortstack{-0.0498\\{\scriptsize [-0.1412, 0.0415]}} & \shortstack{0.0062\\{\scriptsize [-0.0528, 0.0651]}} & \shortstack{0.0775\\{\scriptsize [-0.0017, 0.1566]}} \\
Towns and rural areas ($N=2,792$) & \shortstack{-0.0078\\{\scriptsize [-0.0508, 0.0353]}} & \shortstack{-0.0068\\{\scriptsize [-0.0570, 0.0434]}} & \shortstack{-0.0280\\{\scriptsize [-0.0914, 0.0355]}} & \shortstack{0.0115\\{\scriptsize [-0.0469, 0.0698]}} \\
Interaction test: raw $p$ & 0.056 & 0.112 & 0.606 & 0.015 \\
Interaction test: Holm $p$ & 0.113 & 0.559 & 1.000 & 0.087 \\
\bottomrule\end{tabular}\end{footnotesize}
\notes{Entries compare zero exercise with 60 minutes/day; brackets give pointwise 95\% school-clustered intervals. Each outcome uses a pooled model on 8,247 students, interacting the full exercise block with group membership and sharing other control slopes. Baseline class effects and baseline-score splines match the primary model. Composite units are mean subject z scores; subject columns use subject z scores. Interaction tests have 8 numerator and 111 denominator degrees of freedom. Holm adjustments cover two composite interaction tests and, separately, six subject interaction tests.}
\end{table}

Allowing baseline achievement profiles to vary by gender or school location gives similar results (Appendix Table~\ref{tab:groupscores}). Appendix Table~\ref{tab:groupdetail} reports additional within-group duration contrasts.

\subsection{Robustness and sample selection}

Table~\ref{tab:sensitivity} subjects our baseline findings to a battery of robustness checks. Replacing spline controls for past achievement with linear scores, adding baseline cognitive test controls, or adopting county-level clustering yields similar estimates ($-0.021$ to $-0.024$ mean subject-z units). Restricting the sample strictly to official survey respondents ($N = 8,105$) produces an estimate of $-0.023$ mean subject-z units; applying longitudinal survey weights attenuates the contrast further to $0.0004$ mean subject-z units. Controlling for follow-up classroom fixed effects---which conditions on subsequent class assignment---yields $-0.024$ mean subject-z units.

\begin{table}[!htbp]\centering
\caption{Robustness across specifications and samples}\label{tab:sensitivity}
\begin{footnotesize}\setstretch{1.04}\setlength{\tabcolsep}{4pt}\begin{tabular}{p{3.8cm}rccc r}\toprule
Specification & N & 0 vs. 60 & 30 vs. 60 & 120 vs. 60 & Joint $p$ \\ \midrule
Primary model & 8,247 & \shortstack{-0.0240\\{\scriptsize [-0.0485, 0.0005]}} & \shortstack{-0.0026\\{\scriptsize [-0.0180, 0.0128]}} & \shortstack{-0.0160\\{\scriptsize [-0.0464, 0.0145]}} & 0.301 \\
Linear baseline scores & 8,247 & \shortstack{-0.0244\\{\scriptsize [-0.0488, -0.0001]}} & \shortstack{-0.0021\\{\scriptsize [-0.0175, 0.0133]}} & \shortstack{-0.0155\\{\scriptsize [-0.0458, 0.0149]}} & 0.278 \\
Indicator plus linear time & 8,247 & \shortstack{-0.0181\\{\scriptsize [-0.0386, 0.0024]}} & \shortstack{0.0029\\{\scriptsize [-0.0029, 0.0087]}} & \shortstack{-0.0057\\{\scriptsize [-0.0173, 0.0058]}} & 0.128 \\
Successful follow-ups & 8,105 & \shortstack{-0.0234\\{\scriptsize [-0.0469, 0.0000]}} & \shortstack{-0.0018\\{\scriptsize [-0.0170, 0.0134]}} & \shortstack{-0.0144\\{\scriptsize [-0.0442, 0.0155]}} & 0.304 \\
Panel weighted & 8,105 & \shortstack{0.0004\\{\scriptsize [-0.0307, 0.0316]}} & \shortstack{0.0102\\{\scriptsize [-0.0109, 0.0313]}} & \shortstack{0.0188\\{\scriptsize [-0.0197, 0.0574]}} & 0.537 \\
Follow-up class effects & 8,105 & \shortstack{-0.0242\\{\scriptsize [-0.0480, -0.0004]}} & \shortstack{-0.0006\\{\scriptsize [-0.0161, 0.0149]}} & \shortstack{-0.0143\\{\scriptsize [-0.0448, 0.0162]}} & 0.283 \\
County clustering & 8,247 & \shortstack{-0.0240\\{\scriptsize [-0.0497, 0.0016]}} & \shortstack{-0.0026\\{\scriptsize [-0.0197, 0.0144]}} & \shortstack{-0.0160\\{\scriptsize [-0.0565, 0.0246]}} & 0.301 \\
Add baseline cognition & 8,247 & \shortstack{-0.0210\\{\scriptsize [-0.0455, 0.0034]}} & \shortstack{-0.0023\\{\scriptsize [-0.0178, 0.0132]}} & \shortstack{-0.0156\\{\scriptsize [-0.0456, 0.0144]}} & 0.442 \\
\bottomrule\end{tabular}\end{footnotesize}
\notes{Brackets report pointwise 95\% intervals in mean subject-z units. The successful-follow-up, panel-weighted, and follow-up-class models use the same 8,105 students. County clustering uses 28 counties and $t_{27}$ intervals. Other rows use school clustering. Class effects refer to baseline unless indicated. The linear model tests two exercise terms; spline models test four.}
\end{table}

Appendix Table~\ref{tab:modelcomparison} compares separate-zero linear, quadratic, and alternative-knot exercise functions on the same sample. Their cross-validated prediction errors are close, supporting a flexible examination of curve shape rather than a precise fitted optimum. Appendix Tables~\ref{tab:additional} and~\ref{tab:percent} report exposure ceilings, alternative score normalization, and percentage-of-maximum outcomes. These checks preserve the small adjusted contrasts; some zero-exercise pointwise intervals exclude zero.

As shown in Table~\ref{tab:attrition}, baseline-eligible students without usable follow-up scores exhibit lower baseline test scores ($-0.19$ vs. $0.06$ mean subject-z units) and lower cognitive ability. Outcome availability favors slightly higher-performing students. Across unweighted, survey-weighted, and response-restricted models, the adjusted exercise contrasts remain small.

\begin{table}[!htbp]\centering
\caption{Baseline differences by follow-up outcome availability}\label{tab:attrition}
\begin{small}\setstretch{1.04}\setlength{\tabcolsep}{4pt}\begin{tabular}{p{7.5cm}rrr}\toprule
Baseline variable & Available & Unavailable & Std. difference \\ \midrule
Exercise, minutes/day & 51.29 & 47.56 & 0.066 \\
Baseline achievement index & 0.06 & -0.19 & 0.268 \\
Birth-year-based age & 12.52 & 12.85 & -0.414 \\
Female, percent & 48.32 & 43.35 & 0.100 \\
Agricultural registration, percent & 51.59 & 69.80 & -0.379 \\
Boarding, percent & 29.87 & 38.65 & -0.186 \\
Baseline cognition & 0.07 & -0.26 & 0.402 \\
\bottomrule\end{tabular}\end{small}
\notes{The baseline-eligible cohort includes 8,247 students with valid follow-up outcomes and 639 without. Standardized differences divide the difference in means by the square root of the average group variance.}
\end{table}

\section{Discussion}\label{sec:discussion}

Our results place academic history at the very center of adolescent time-use research. Students who report different exercise regimes systematically differ in baseline educational capital. Once these baseline differences are accounted for, the zero-versus-one-hour achievement contrast attenuates by 85.1 percent. Baseline test scores summarize earlier educational inputs, ability, parental investments, and potentially prior exercise. Conditioning on this history compares students with similar observed academic records and substantially changes the estimated exercise gradient. The estimates describe conditional associations with subsequent achievement \citep{hernan2020}.

Methodologically, our framework separates two comparisons combined in a continuous quadratic specification: the gap between zero exercise and a positive reference duration, and the gradient across positive durations. The zero-exercise contrast is negative, while the reported positive-duration contrasts are small and the curvature test does not reject linearity. Separating these margins makes the interpretation of a fitted turning point explicit: an intermediate academic optimum must be supported by the positive-duration profile itself.

The measurement structure also shapes the research design. Self-reported exercise over a single reference week is an imperfect proxy for long-term physical activity, and survey records lack information on exercise intensity and specific displaced activities. Individual fixed-effects models remove time-invariant student characteristics, but the changed exercise instrument prevents interpreting the two reports as a consistently measured change in activity. The prospective specification uses baseline exercise to study subsequent achievement conditional on observed academic history and classroom environments.

Finally, our dependent variable captures relative within-school exam performance rather than absolute learning gains on a standardized national metric. Future research utilizing harmonized multi-wave assessments and objective activity trackers (such as accelerometers) would provide deeper insight into how structured physical activity programs interact with dynamic human capital accumulation.

\section{Conclusion}

This paper provides a longitudinal examination of the relationship between adolescent extracurricular exercise and academic achievement. Using a two-wave prospective panel from the CEPS, we show that controlling flexibly for baseline academic achievement reduces the zero-versus-one-hour achievement gap by 85.1 percent, with a joint exercise-test $p$-value of 0.301 and a curvature-test $p$-value of 0.659. Reported positive-duration contrasts are small, gender profiles are similar, and evidence of school-location variation is strongest for English.

These findings sharpen the interpretation of exercise evidence in educational policy. Cross-sectional correlations and fitted polynomial turning points should not be treated as behavioral production functions or prescriptive daily targets. The 85.1 percent attenuation makes prior academic performance central to that assessment. Claims about examination gains from particular activity programs require evidence on those programs, rather than an optimum inferred from students' existing time allocations.

\section*{Data and Code Availability}

The study uses the CEPS 2013--2014 baseline and 2014--2015 follow-up releases. Replication materials document variable construction and estimation. Access to the underlying microdata is subject to CEPS authorization.

\clearpage\appendix
\section{Variable Construction and Sample Checks}\label{app:variables}

\paragraph{Linkage and response.}
Student identifiers are unique in both waves. Parent records merge one-to-one by student, and principal records merge many-to-one by school. The follow-up student file contains 10,750 records: 9,449 successful follow-ups, 830 unsuccessful follow-ups, and 471 new entrants. All 10,279 baseline seventh-graders have matching records; the successful survey-follow-up rate is 91.93 percent. The analytic cohort excludes new entrants because they lack baseline exposure.

Among unsuccessful follow-ups, 176 have complete, valid three-subject scores; 142 also satisfy baseline eligibility. These 142 comprise 90 classified as transferred, 22 not currently in school, 16 still at their original school, 12 of unknown type, and two recorded as deceased. Status dates relative to the examination are unavailable. Normalization and clustering use the retained baseline school identifiers, and the respondent-only sensitivity excludes these records. Follow-up class identifiers are available for the 8,105 successful follow-ups in the analytic sample.

\paragraph{Exercise and controls.}
Baseline exercise uses \texttt{b15c1}, \texttt{b15c2}, \texttt{b16c1}, and \texttt{b16c2}; all four components must be observed \citep{cepsquestionnaire}. Follow-up exercise frequency and duration are \texttt{w2c13a} and \texttt{w2c13b}. These definitions follow the supplied questionnaire and crosswalk (\citealp{ceps2015q}; \citealp{cepscrosswalk}).

Age equals 2013 minus birth year. Parental education is grouped into primary or less, junior secondary, senior-secondary or vocational equivalents, and college or above. Household economic status enters categorically. Recalled sixth-grade difficulty enters separately by subject, with ``subject not studied'' treated as missing. Only children have zero siblings; the number of siblings is top-coded at five. The primary complete-case rule requires all included controls. Follow-up family and health variables are outside the baseline adjustment set.

\paragraph{Examination scores.}
The follow-up raw marks are \texttt{w2chn}, \texttt{w2mat}, and \texttt{w2eng}; the corresponding \texttt{w2up} fields give maximum marks \citep{ceps2017}. A valid mark lies between zero and a positive recorded maximum. There are 28 mathematics and five English marks above their maxima; the violations affect 32 records with all three marks. Maximum marks are constant within each recorded school and subject.

Subject normalization uses all valid follow-up records, including new entrants and unsuccessful survey follow-ups with school scores. Each subject has its own reference mean and sample standard deviation, computed before complete-case and exposure restrictions. The outcome requires all three standardized subjects. One sensitivity excludes new entrants from the reference; another includes nonnegative above-maximum scores and recomputes normalization. The percentage-of-maximum outcome averages $100R_{isj,1}/\mathrm{Max}_{sj,1}$ across subjects.

\paragraph{Timing and adjustment.}
The follow-up examination is the autumn-2014 midterm \citep{ceps2017}. Baseline interviews took place in the 2013--2014 school year. Their exact ordering relative to baseline examinations is not established, and the fieldwork manual allows replacement examinations when intended results are unavailable \citep{ceps2014field}. Baseline achievement is therefore treated as observed academic history. The primary score functions use each subject's main-sample 10th, 50th, and 90th percentiles as spline knots. Main, subject, and group comparisons share these knots; sensitivity analyses that change the estimation sample recompute the baseline-score knots within that sample. Exercise knots remain fixed at the baseline-eligible seventh-grade reference percentiles. Follow-up class effects condition on subsequent class assignment and are reported as a sensitivity analysis.

\section{Spline Basis and Clustered Inference}\label{app:inference}

For ordered knots $k_1<\cdots<k_K$, the spline representation contains $T$ and
\begin{equation}
h_j(T)=\frac{(T-k_j)_+^3-\frac{k_K-k_j}{k_K-k_{K-1}}(T-k_{K-1})_+^3+\frac{k_{K-1}-k_j}{k_K-k_{K-1}}(T-k_K)_+^3}{(k_K-k_1)^2},
\quad j=1,\ldots,K-2,
\end{equation}
where $z_+=\max(z,0)$. The four-knot exposure block is $D,T,h_1(T),h_2(T)$. Reparameterizing the positive-exercise coefficient yields the centered specification in equation~\eqref{eq:main}. The reported zero-versus-60-minute contrast combines the indicator and duration coefficients.

After removing baseline-class means, let $\widetilde X$ and $\widetilde Y$ denote the design and outcome, $B=(\widetilde X'\widetilde X)^{-1}$, and $S_g=\widetilde X_g'\widehat\varepsilon_g$. The school-cluster covariance is
\begin{equation}
\widehat V=qB\left(\sum_{g=1}^{G}S_gS_g'\right)B,
\qquad q=\frac{G}{G-1}\frac{N-1}{N-C-p}.
\end{equation}
The primary model has $N=8,247$, $G=112$, $C=221$, and within-design rank $p=34$. Weighted regressions use weighted class means and the corresponding transformed design. Pointwise intervals use $t_{G-1}$ critical values; joint tests use an $F$ reference with denominator degrees of freedom $G-1$. County clustering uses 28 groups. Independent explicit-dummy OLS and WLS implementations verify the coefficients and clustered exposure covariance.

For simultaneous inference, draw independent $w_g\in\{-1,1\}$ for schools and form
\begin{equation}
\Delta\beta^*=\sqrt q\,B\sum_{g=1}^{G}w_gS_g.
\end{equation}
Each draw takes the maximum of $|L_t\Delta\beta^*|/\widehat{\mathrm{se}}(L_t\widehat\beta)$ over integer-minute durations $t=5,\ldots,120$, excluding the reference at 60. The 95th percentile of 20,000 draws is 2.437 (seed 2026101202). The simulation conditions on the fitted design and score normalization. The simultaneous family covers the stated positive-duration grid; the zero-exercise contrast and subject-specific comparisons have separate pointwise intervals.

\paragraph{Subject and group comparisons.}
Subject comparisons retain the common design and account for correlated outcomes through cross-subject school score products. If $S_{gj}$ is the score vector for subject $j$, the cross-equation covariance block is $qB(\sum_g S_{gj}S_{gk}')B$. An eight-restriction Wald test evaluates whether all four exercise coefficients agree across the three subjects. Gender and location models interact the four exercise terms with group indicators while sharing the other control slopes. Their joint tests have four and eight numerator degrees of freedom, respectively. Holm adjustments cover two composite interaction tests and, separately, six subject interaction tests. Individual group contrasts retain pointwise intervals.

\section{Additional Results}

\begin{table}[!htbp]\centering
\caption{Exposure, scoring, and fieldwork sensitivity}\label{tab:additional}
\begin{footnotesize}\setstretch{1.04}\setlength{\tabcolsep}{4pt}\begin{tabular}{p{3.8cm}rccc r}\toprule
Specification & N & 0 vs. 60 & 30 vs. 60 & 120 vs. 60 & Joint $p$ \\ \midrule
Exercise cap: 3 h & 7,950 & \shortstack{-0.0273\\{\scriptsize [-0.0519, -0.0026]}} & \shortstack{-0.0077\\{\scriptsize [-0.0257, 0.0103]}} & \shortstack{-0.0167\\{\scriptsize [-0.0477, 0.0142]}} & 0.060 \\
Exercise cap: 4 h & 8,149 & \shortstack{-0.0253\\{\scriptsize [-0.0498, -0.0007]}} & \shortstack{-0.0042\\{\scriptsize [-0.0201, 0.0117]}} & \shortstack{-0.0154\\{\scriptsize [-0.0456, 0.0149]}} & 0.137 \\
Exercise cap: 8 h & 8,298 & \shortstack{-0.0238\\{\scriptsize [-0.0484, 0.0008]}} & \shortstack{-0.0025\\{\scriptsize [-0.0181, 0.0132]}} & \shortstack{-0.0179\\{\scriptsize [-0.0469, 0.0111]}} & 0.296 \\
Retain above-max scores & 8,277 & \shortstack{-0.0253\\{\scriptsize [-0.0498, -0.0009]}} & \shortstack{-0.0038\\{\scriptsize [-0.0192, 0.0116]}} & \shortstack{-0.0178\\{\scriptsize [-0.0484, 0.0128]}} & 0.266 \\
Tracked-cohort reference & 8,247 & \shortstack{-0.0239\\{\scriptsize [-0.0483, 0.0006]}} & \shortstack{-0.0026\\{\scriptsize [-0.0180, 0.0128]}} & \shortstack{-0.0158\\{\scriptsize [-0.0464, 0.0148]}} & 0.306 \\
Baseline autumn sample & 5,026 & \shortstack{-0.0235\\{\scriptsize [-0.0522, 0.0052]}} & \shortstack{0.0022\\{\scriptsize [-0.0180, 0.0224]}} & \shortstack{0.0015\\{\scriptsize [-0.0421, 0.0451]}} & 0.260 \\
Baseline spring sample & 3,221 & \shortstack{-0.0229\\{\scriptsize [-0.0684, 0.0225]}} & \shortstack{-0.0129\\{\scriptsize [-0.0367, 0.0109]}} & \shortstack{-0.0429\\{\scriptsize [-0.0823, -0.0035]}} & 0.248 \\
\bottomrule\end{tabular}\end{footnotesize}
\notes{Brackets report pointwise 95\% intervals in mean subject-z units. Exercise knots remain fixed; baseline-score knots use each estimation sample. Retaining above-maximum scores recomputes normalization; the tracked-cohort reference excludes new entrants. Season samples differ in school composition and interview timing.}
\end{table}

\begin{table}[!htbp]\centering
\caption{Alternative outcome: mean percentage of subject maximum marks}\label{tab:percent}
\begin{small}\setstretch{1.04}\setlength{\tabcolsep}{4pt}\begin{tabular}{rrr}\toprule
Minutes/day & Estimate & Pointwise 95\% CI \\ \midrule
0 & -0.3766 & [-0.8897, 0.1366] \\
30 & -0.0434 & [-0.3726, 0.2858] \\
90 & -0.1524 & [-0.5547, 0.2499] \\
120 & -0.2571 & [-0.8179, 0.3037] \\
\bottomrule\end{tabular}\end{small}
\notes{N=8,247. The outcome averages 100 times raw marks divided by subject maximum marks. Units are percentage-of-maximum points. The model otherwise matches the primary specification. The joint exercise-test p value is 0.495.}
\end{table}

\begin{table}[!htbp]\centering
\caption{Functional-form and covariate comparisons}\label{tab:modelcomparison}
\begin{footnotesize}\setstretch{1.04}\setlength{\tabcolsep}{4pt}\begin{tabular}{p{6.1cm}crr}\toprule
Specification & 0 vs. 60 minutes & Exercise joint $p$ & Held-out RMSE \\ \midrule
Primary model & \shortstack{-0.0240\\{\scriptsize [-0.0485, 0.0005]}} & 0.301 & 0.432127 \\
Linear baseline scores & \shortstack{-0.0244\\{\scriptsize [-0.0488, -0.0001]}} & 0.278 & 0.433649 \\
Linear exercise duration & \shortstack{-0.0181\\{\scriptsize [-0.0386, 0.0024]}} & 0.128 & 0.432026 \\
Quadratic exercise duration & \shortstack{-0.0178\\{\scriptsize [-0.0394, 0.0039]}} & 0.225 & 0.432064 \\
Five exercise knots & \shortstack{-0.0332\\{\scriptsize [-0.0640, -0.0025]}} & 0.267 & 0.432153 \\
Add baseline cognition & \shortstack{-0.0210\\{\scriptsize [-0.0455, 0.0034]}} & 0.442 & 0.431066 \\
Omit boarding and difficulty & \shortstack{-0.0276\\{\scriptsize [-0.0521, -0.0030]}} & 0.201 & 0.432840 \\
\bottomrule\end{tabular}\end{footnotesize}
\notes{All models use the same 8,247 students and baseline class fixed effects. The primary model uses four exercise knots and three knots for each baseline subject score. Alternative exercise forms retain a separate participation indicator and the same score adjustment. Brackets give pointwise 95\% school-clustered intervals. RMSE pools five repetitions of five-fold school-held-out validation of class-demeaned scores. Baseline-score knots are recomputed in each training fold. School score normalization and exercise knots are held fixed. The diagnostic evaluates within-class score differences; smaller values indicate better prediction on this scale.}
\end{table}

\begin{table}[!htbp]\centering
\caption{Composite-achievement contrasts within gender and location groups}\label{tab:groupdetail}
\begin{footnotesize}\setstretch{1.04}\setlength{\tabcolsep}{4pt}\begin{tabular}{p{4.0cm}cccc}\toprule
Group & 0 vs. 60 & 30 vs. 60 & 90 vs. 60 & 120 vs. 60 \\ \midrule
Boys & \shortstack{-0.0277\\{\scriptsize [-0.0662, 0.0108]}} & \shortstack{-0.0089\\{\scriptsize [-0.0335, 0.0158]}} & \shortstack{-0.0138\\{\scriptsize [-0.0434, 0.0158]}} & \shortstack{-0.0197\\{\scriptsize [-0.0609, 0.0214]}} \\
Girls & \shortstack{-0.0214\\{\scriptsize [-0.0503, 0.0076]}} & \shortstack{0.0014\\{\scriptsize [-0.0168, 0.0195]}} & \shortstack{-0.0036\\{\scriptsize [-0.0335, 0.0262]}} & \shortstack{-0.0133\\{\scriptsize [-0.0533, 0.0267]}} \\
City or county center & \shortstack{-0.0553\\{\scriptsize [-0.0863, -0.0244]}} & \shortstack{-0.0199\\{\scriptsize [-0.0426, 0.0029]}} & \shortstack{-0.0478\\{\scriptsize [-0.0765, -0.0191]}} & \shortstack{-0.0684\\{\scriptsize [-0.1073, -0.0296]}} \\
Urban / rural--urban fringe & \shortstack{0.0113\\{\scriptsize [-0.0520, 0.0745]}} & \shortstack{0.0141\\{\scriptsize [-0.0143, 0.0424]}} & \shortstack{-0.0013\\{\scriptsize [-0.0436, 0.0410]}} & \shortstack{-0.0029\\{\scriptsize [-0.0615, 0.0558]}} \\
Towns and rural areas & \shortstack{-0.0078\\{\scriptsize [-0.0508, 0.0353]}} & \shortstack{0.0117\\{\scriptsize [-0.0185, 0.0419]}} & \shortstack{0.0312\\{\scriptsize [-0.0090, 0.0715]}} & \shortstack{0.0402\\{\scriptsize [-0.0159, 0.0963]}} \\
\bottomrule\end{tabular}\end{footnotesize}
\notes{The reference is 60 minutes/day. Estimates and pointwise 95\% intervals come from the pooled interaction models in Tables~\ref{tab:gender} and~\ref{tab:location}, with school-clustered inference on 112 schools. Units are mean subject z scores.}
\end{table}

\begin{table}[!htbp]\centering
\caption{Group comparisons with flexible achievement persistence}\label{tab:groupscores}
\begin{footnotesize}\setstretch{1.04}\setlength{\tabcolsep}{4pt}\begin{tabular}{p{3.7cm}rrrr}\toprule
Outcome / comparison & \shortstack{Common score\\slopes: raw $p$} & \shortstack{Common score\\slopes: Holm $p$} & \shortstack{Group-specific\\slopes: raw $p$} & \shortstack{Group-specific\\slopes: Holm $p$} \\ \midrule
Composite / gender & 0.670 & 0.670 & 0.597 & 0.597 \\
Composite / location & 0.056 & 0.113 & 0.072 & 0.143 \\
Chinese / gender & 0.686 & 1.000 & 0.462 & 1.000 \\
Chinese / location & 0.112 & 0.559 & 0.113 & 0.567 \\
Mathematics / gender & 0.482 & 1.000 & 0.501 & 1.000 \\
Mathematics / location & 0.606 & 1.000 & 0.639 & 1.000 \\
English / gender & 0.546 & 1.000 & 0.574 & 1.000 \\
English / location & 0.015 & 0.087 & 0.023 & 0.137 \\
\bottomrule\end{tabular}\end{footnotesize}
\notes{All entries test the exercise-by-group interactions. The alternative model also interacts the six baseline-score spline terms with group indicators, adding six terms for gender and twelve for location. Other controls remain common. Both models use the same 8,247 students, fixed knots, baseline class effects, and school-clustered inference. Holm adjustments use the same two composite and six subject tests in each specification.}
\end{table}

\FloatBarrier
\clearpage
\section{Reproduction}

The analysis uses four input files:
\begin{itemize}
\item Baseline student records: \texttt{cepsw1studentEN.dta}.
\item Baseline parent records: \texttt{cepsw1parentEN.dta}.
\item Baseline principal records: \texttt{cepsw1principalEN.dta}.
\item Follow-up student records: \texttt{cepsw2studentEN.tab}.
\end{itemize}
Baseline data preparation is implemented in \texttt{audit\_models.py}. The program \texttt{panel\_analysis.py} links the cohort, constructs outcomes, estimates the models, and generates the simultaneous band. Separate scripts audit record linkage and validate the numerical results against regressions with explicit class indicators.

The source package contains aggregate outputs and a self-contained LaTeX manuscript with embedded tables, figure coordinates, and references. Input hashes identify the release used in estimation. Rerunning the analysis requires authorized copies of the microdata.
\end{document}